**Opening long-time investigation window of living matter by nonbleaching phase intensity nanoscope: PINE**

*Guangjie Cui[1#], Yunbo Liu[1#], Di Zu[1], Xintao Zhao[1], Zhijia Zhang[1], Do Young Kim[1], Pramith Senaratne[1], Aaron Fox[1], David Sept[2], Younggeun Park[3], Somin Eunice Lee[1]\**

[1]Department of Electrical & Computer Engineering, Biomedical Engineering, Applied Physics, Biointerfaces Institute, Macromolecular Science & Engineering, University of Michigan, [2]Department of Biomedical Engineering, University of Michigan, [3]Department of Mechanical Engineering, University of Michigan
[#]Authors equally contributed. * To whom correspondence should be addressed.

## Abstract

Fundamental to all living organisms and living soft matter are emergent processes in which the reorganization of individual constituents at the nanoscale drives group-level movements and shape changes at the macroscale over time. However, light-induced degradation of fluorophores, photobleaching, is a significant problem in extended live cell imaging in life science, and excellent prior arts of STED, MINSTED, or MINFLUX intentionally apply photobleaching optics for super-resolution, yet lack extended live-cell imaging capability as fluorophores reach an irreversible photobleaching limit. Here, we report an innovative way of opening a long-time investigation window of living matter by a nonbleaching **p**hase **i**ntensity **n**anoscop**e**: PINE. We accomplish phase-intensity separation to obtain phase differences between electric field components, such that nanoprobes within a diffraction limited region are distinguished from one another by an integrated phase-intensity multilayer thin film (polyvinyl alcohol/liquid crystal). We obtain distributional patterns of precisely localized nanoprobes to overcome a physical limit to resolve sub-10 nm cellular architectures, and achieve the first dynamic imaging of nanoscopic reorganization over 250 hours using nonbleaching PINE. Since PINE allows a long-time investigation window, we discover nanoscopic rearrangements synchronized with the emergence of group-level movements and shape changes at the macroscale according to a set of interaction rules. We believe that PINE and its application presented here provide a mechanistic understanding of emergent dynamics important in cellular and soft matter reorganization, self-organization, and pattern formation.

**Introduction**

How macroscale groups emerge over time from their individual constituents is fundamental to self-assembly, self-organization and pattern formation in material science[1] and biology[2,3]. The diverse range of emergent dynamics[4–6] spans at least three orders of magnitude in length scales and at least five orders of magnitude in time scales. Fluorescence super-resolution[7–11] – in which STED, MINSTED, MINFLUX intentionally apply photobleaching and STORM, dSTORM, PALM, dPALM cycle fluorescence states before photobleaching - have unlocked optical imaging across the spatial domain down to the sub-10 nm length scale. However, light-induced degradation of fluorophores is permanent which ultimately sets an irreversible photobleaching limit and restricts observation time, making long-time super-resolution in live cells difficult to achieve. Thus, a long-time super-resolution method that can open a long-time investigation window down to the sub-10 nm length scale is needed for visualizing emergent nanoscale-to-macroscale dynamics but has not been yet reported.

Nanoprobes undergoing elastic scattering processes without photobleaching offers a nonbleaching approach to achieve long-time super-resolution. However, current methods[12–16] by moving analyzers/polarizers are subject to displacements and imprecise localization, making precise super-resolution difficult. Current demonstrations have resolved only two to few nanoprobes, whereas populations (thousands) are needed to form patterns of underlying cellular architectures. Additionally, all of the aforementioned methods have not achieved super-resolution down to the sub-10 nm length scale.

Herein, we present a nonbleaching **p**hase **i**ntensity **n**anoscop**e**, PINE to open a long-time investigation window of living matter down to the sub-10 nm length scale. We designed an integrated phase-intensity multilayer thin film of polyvinyl alcohol and liquid crystalline polymers (herein referred to as PI) (Figure 1a), enabling populations of randomly distributed nanoprobes (gold nanorods) to exhibit phase differences between electric field components in a stochastic manner (Figure 1b, 1c). Owing to the tunability of PI, we achieved phase-intensity separation of nanoprobes within a diffraction limited region by modulating phase differences between electric field components. As PI is displacement-free, precise localization of nanoprobe populations (thousands) form patterns of underlying cellular architectures enabling precise super-resolution down to the sub-10 nm length scale. We quantified distributional parameters to overcome a physical limit to resolve sub-10 nm cellular architectures. Using nonbleaching PINE, we obtained the first dynamic imaging of nanoscale reorganization over 250 hours, which outperforms state-of-the-art fluorescence super-resolution by more than two orders of magnitude. With a long-time investigation window by PINE (Figure 1d), we identified emergent **sy**nchronized **in**dividual-group **c**ontraction-expansion, SYNC, in which nanoscopic rearrangements coordinated with the emergence of group-level movements and shape changes at the macroscale during the cell division process according to a set of interaction rules. We are able to achieve long-time sub-10 nm nanoscopy and follow the emergence of large scale collectives involving hundreds of cellular architectures (> 900) in cells, all of which cannot be obtained using existing fluorescence[7–11] or nonbleaching[12–16] methods.

## Results

**Principle of PI device design**

We fabricated PI (Figure 2a) for precise control of phase differences between electric field components to accomplish phase-intensity separation of nanoprobes within a diffraction limited region (Figure 2b). The compact multilayer design (~600 μm thickness) was integrated into the infinity space between the objective and tube lens (Figure 1c), enabling high transmission for nanoscopy. Scattering[17–23] without photobleaching was used. Input light to PI was the scattered light from a population of nanoprobes (gold nanorods). Scattered light was reshaped, phase modulated, and converted from phase to intensity by PI. Phase differences between electric field components enable nanoprobes within a diffraction limited region to be distinguished from one another by phase-intensity separation. Because the optical axis is stationary, PI is displacement-free for precise nanoscopy. Thus,

$$\vec{E} = \frac{1}{2}\begin{bmatrix} 1 & 0 \\ 0 & 0 \end{bmatrix} \begin{bmatrix} e^{i\frac{\gamma}{2}}+e^{-i\frac{\gamma}{2}} & e^{i\frac{\gamma}{2}}-e^{-i\frac{\gamma}{2}} \\ e^{i\frac{\gamma}{2}}+e^{-i\frac{\gamma}{2}} & e^{i\frac{\gamma}{2}}-e^{-i\frac{\gamma}{2}} \end{bmatrix} \begin{bmatrix} e^{i\delta_x} & 0 \\ 0 & e^{i\delta_y} \end{bmatrix} \begin{bmatrix} E_x \\ E_y \end{bmatrix} \qquad (1)$$

where $\delta = \delta_y - \delta_x$ is the phase difference between the scattered electric field components $E_x$ and $E_y$, $\gamma = \int_0^D \frac{n_e n_o dz}{\sqrt{n_e^2 \cos^2\theta + n_o^2 \sin^2\theta}} - n_o D$ is the phase retardation, and $D$, $n_o$, $n_e$, and $\theta$ are the layer thickness, ordinary refractive index, extraordinary refractive index, and molecular alignment with respect to the applied electric field. The Jones matrices represent a polyvinyl alcohol layer designed as a fixed retarding element for reshaping, 4-cyano-4′-pentylbiphenyl and 4-(3-acryloyoxypropyloxy) benzoic acid 2-methyl-1,4-phenylene ester layer designed as a variable retarding element for phase modulation, and polyvinyl alcohol-iodine layer designed as a fixed linear polarizing element for phase-to-intensity conversion. The resulting variation of intensity corresponded to subsets of nanoprobes where

$$I(\gamma) = \cos^2\left(\frac{\gamma}{2}\right) E_x^2 + \sin^2\left(\frac{\gamma}{2}\right) E_y^2 + 2\cos\left(\frac{\gamma}{2}\right) E_x \sin\left(\frac{\gamma}{2}\right) E_y \sin(\delta) \qquad (2)$$

PI was fabricated by a deposition process, where precursor of 4-(3-acryloyoxypropyloxy) benzoic acid 2-methyl-1,4-phenylene ester, 1-hydroxycyclohexyl phenyl ketone, and 4-cyano-4′-pentylbiphenyl was deposited and cured between two indium tin oxide-polyethylene terephthalate substrates on top of polyvinyl alcohol-iodine layer. Polyvinyl alcohol was layered on top of the aforementioned layers. We characterized the material alignment uniformity and millisecond response time. During phase-intensity modulation, the center coordinates were expressed as

$$x_\gamma = x_0,$$
$$y_\gamma = y_0 \qquad (3)$$

where $x_0$ and $y_0$ are the center coordinates prior to modulation. Here, the optical axis was stationary such that $(x_\gamma, y_\gamma)$ is equal to $(x_0, y_0)$, enabling zero displacement for precise nanoscopy. This allows for precise localization and eliminates potential error propagation in scaling up[24]. In scaling up, populations (thousands) of precisely localized nanoprobes forms patterns of underlying architectures of arbitrary shape. Conventional microscopy blurs the distribution of nanoprobes convolved with the point spread function (PSF) of the imaging system. To obtain sub-10 nm information, nanoprobes within a diffraction limited region were isolated by phase-intensity separation with zero displacement for precise nanoscopy such that the distribution of precisely localized nanoprobes forms pattern of underlying architectures. From the distribution of precisely localized nanoprobes, surface or curvilinear features *f(p)* were defined. By defining a new parameter $\chi$, sub-10 nm information was obtained from the distribution $\chi(p)$.

**Sub-10 nm PINE**

To validate the phase-intensity separation concept, we investigated nanoprobes (gold nanorods) fabricated by electron beam lithography in order to systematically vary geometrical parameters. For this validation study, we chose electron beam lithography over chemical synthesis because of precise fabrication control of geometrical parameters. As a model, we fabricated plasmonic gold nanorods to determine whether phase differences between electric field components could be used to segregate nanorods in a diffraction limited region. Due to their high aspect ratio geometries significantly smaller than the wavelength, conduction electrons collectively oscillate in phase, resulting in strong scattering cross-sections[25,26,35–38,27–34]. As an initial test, the scattered light from a pair of nanorods was reshaped by PI. When PI modulated $\gamma$ from 0 to $2\pi$ rad, the phase difference between $E_x$ and $E_y$ was varied from $-\pi$ to $\pi$ rad and the positions of each nanorod emerged. PINE precisely determined the spatial positions of each gold nanorod (Figure 2b and 2c). Localization was firstly optimized based on geometrical parameter, aspect ratio, *via* simulation, whereby we then selected aspect ratio 2.9 to be fabricated by electron beam lithography. Due to the delicate nature of electron beam fabricated samples, a long working distance low NA objective and dry condenser with a diffraction limit of 450 nm were used. In Figure 2d, we benchmarked geometrical parameter $\ell$ imaged by PINE against $\ell$ measured by scanning electron microscopy. We fixed $\alpha$ and then varied $\ell$ between 100 nm to 400 nm. We validated localization accuracy in agreement with scanning electron microscopy measurements with an $R^2$ of 0.97 (Figure 2d). We then turned to geometrical parameter $\alpha$. In Figure 2e, we varied $\alpha$ between 2° to 100° at a fixed known subdiffraction $\ell$ of 300 nm which was below the diffraction limit of 450 nm. Owing to the displacement-free multilayer, nearly identical nanorods as $\alpha \to 0$ could be precisely localized (red curve Figure 2e). The minimum $\alpha$ that two proximal nanoprobes will require at an effective subdiffraction $\ell$ is 2°. To answer the question of the minimum $\alpha$ that two nanoprobes at the minimum $\ell$ will require, we then varied both $\alpha$ and $\ell$. The minimum $\alpha$ that two nanoprobes at the minimum $\ell$ of 80 nm is 45°. Smaller $\alpha$ are resolvable at larger subdiffraction $\ell$ below the diffraction limit. We verified multiple nanoprobes within a diffraction limited region were phase-intensity separated by PI (Figure 2f). We estimate approximately twenty nanorods can be resolved in a diffraction limited spot.

    Randomly distributed nanoprobes introduce a stochastic variation within a population (Figure 3a). We reasoned their phase differences between electric field components should be also stochastically distributed. We utilized this stochasticity for the purpose of enhancing resolution. In order to form patterns of underlying cellular architectures, structurally stabilized nanoprobes in which scaling up to distributions of nanoprobes are required. Traditionally, gold nanorods[39,40] are synthesized with a strong cationic charge from hexadecyltrimethylammonium bromide capping in which achieving 100% CTA+ free necessary for structural stability has been

historically difficult due to the strong cationic charge[41]. Here, we chemically synthesized with bromide-free surfactants to yield 100% CTA+ free monodisperse gold nanorods with size variability of ~5%.[42,43] We achieved structurally stabilized distributions of nanoprobes for nanoscopic imaging (Figure 3). In scaling up, populations of precisely localized nanoprobes formed patterns of the underlying cellular architectures. In cells, actin plays key roles both in bundled and single filament forms. Actin was labeled with CTA+ free nanorods conjugated with antibodies targeting actin. Scanning electron microscopy images verified nanoprobes were randomly distributed along actin bundles. PI reshaped scattered light from the randomly distributed population of nanoprobes. PI then modulated $\gamma$ from 0 to $2\pi$ rad (Figure 3b) which varied the phase difference between $E_x$ and $E_y$ from $-\pi$ to $\pi$ rad. After phase-to-intensity conversion, the resulting variation of intensity corresponded to subsets of precisely localized nanoprobes. To form patterns, we were able to scalably resolve nanoprobe populations from few to 7,769 nanoprobes (up to ~$10^2/\mu m^2$). Scaling up allowed for distributions of precisely localized nanoprobes to form patterns of the underlying actin bundles in agreement with scanning electron microscopy. Underlying architectures were identified for nanoprobe labeling densities of ~$10^2/\mu m^2$ based on distribution of distances from every nanoprobe to its first neighbor following a distribution function[44], enabling continuous features to be defined from the discrete nanoprobe distributions in agreement with scanning electron microscopy.

In contrast to bundles we observed earlier, actin is often in single filament form in cells[45]. In SH-SY5Y cells, nanoprobes within a diffraction limited region were isolated by phase-intensity separation by modulating $\gamma$ from 0 to $2\pi$ rad using PI. Precisely localized nanoprobes formed patterns of the underlying single filaments. While the minimum $\ell$ is 80 nm, sub-10 nm information can be obtained. Surface or curvilinear features $f(p)$ were defined. Parameter $\chi$ was obtained from the pattern of precisely localized nanoprobe populations and $f(p)$. From the distribution $\chi(p)$, PINE revealed actin width to be 8.0 nm (Figure 3c) consistent with width of actin filaments in cells[46].

Notably, sub-10 nm features revealed single filament architectures (Figure 3c iii) and branched architectures (Figure 3c iv) which were indistinguishable from each other in the absence of sub-10 nm information. Both controls - absence of both nanoprobes and antibody; presence of nanoprobes without antibody - were negative. To our knowledge, this is the first sub-10 nm demonstration by a nonbleaching nanoscopy method owing to the displacement-free mechanism of PI for precise nanoscopy.

**Long-time PINE (Model: Emergent dynamics SYNC)**

A longstanding question is how macroscale groups emerge from their individual constituents. Commonalities between diverse systems suggest general rules[47–50] guiding the collective reorganization of individuals into groups. This has prompted models of emergent phenomena ranging from microscale forces[51–53] and cellular forces at the macroscale[54,55] to microscale chemical reactions[56,57] and macroscale reactions among networks[58], underscoring a need for experiments and theories bridging across diverse length scales and time scales. Using PINE, the ability to bridge across diverse length scales and time scales provides a unique opportunity to shed light on the collective reorganization of individuals and groups (Figure 1b iii).

When large scale collectives involve hundreds of individual constituents, increasing experimental evidence[5,59] suggests that collective behavior may arise from local interactions versus global instructions. From a set of local interaction rules, we describe **sy**nchronized i**n**dividual-group **c**ontraction-expansion, SYNC, in which individual constituents coordinate with group level movements and shape changes at the macroscale (Figure 4a). A group of $N$ individuals will expand or contract based on connectivity between neighboring individuals. Connectivity $c$ is defined as the ability of individuals to connect with one or more others. **Rule 1**: An individual will occupy more short-range space to increase connectivity with neighbors.

$$s_i = \sum_{k \neq i} \frac{p_i(t) - p_k(t)}{|p_i(t) - p_k(t)|} \tag{4}$$

where $s_i$ is the direction of movement of individual *i* within a short-range space $\rho$, $p_i$ is the position vector of an individual *i*, and $p_k$ is the position vector of a neighbor *k* within the short-range space $\rho$. This simulates the movement direction of an individual to alter connectivity with short-range neighbors. **Rule 2**: Long-range space between individuals will decrease to increase connectivity with neighbors.

$$l_i = -\sum_{j \neq i} \frac{p_i(t) - p_j(t)}{|p_i(t) - p_j(t)|} \tag{5}$$

where $l_i$ is the direction of movement of individual *i* within a long-range space $P$, $p_i$ is the position vector of an individual *i*, and $p_j$ is the position vector of a neighbor *j* within a long-range space $P$. This represents the movement direction of an individual to alter connectivity with long-range neighbors. **Rule 3**: Group contracts with increased connectivity. Group expands with decreased connectivity.

$$M_i(t + \Delta t) = -\sum_{i \neq j} \frac{c[p_i(t) - p_j(t)] - s_i}{|c[p_i(t) - p_j(t)] - s_i|} \tag{6}$$

where $M_i$ is the direction of movement of individual *i* over time *t* influenced by connectivity *c*. This manifests in expansion or contraction of the group on the basis of connectivity between neighboring individuals. Thus, *c* was increased for expansion and *c* was decreased for contraction. SYNC was quantified by resultant connection defined as $(1 - C)$. We found there was a threshold group size for collective behavior at the group level. At the threshold group size, expansion-contraction behavior commenced (Figure 4e i), meaning sufficient interactions between individuals and neighboring individuals were necessary. Thereafter, as the group size became larger, the expansion-contraction behavior increased. Threshold group size may impact expansile and contractile forces of a connected actin network (group) influenced by filament length and density[54].

**Long-time PINE (Experiment: Emergent dynamics SYNC)**

We applied the model together with PINE to investigate the emergent process of cell division[60]. In order to investigate nanoscopic dynamics, we benchmarked the temporal capabilities of PINE against fluorescence super-resolution (Figure 4b). Benchmarking nanoprobe-labeled actin bundles resolved by PINE with fluorophore-labeled actin bundles resolved by fluorescence super-resolution, subdiffraction resolution was preserved under continuous analysis using PINE, whereas with fluorescence super-resolution, subdiffraction capabilities were lost over time (<200 s) with continuous exposure to illumination. To validate long-time observation capabilities, we followed the temporal evolution of cellular architectures (actin) (Figure 4c) and observed nanoscale rearrangements below the diffraction limit in agreement with theoretical modeling. PINE can uniquely reveal individual constituents at the sub-10 nm scale, their meso- and macroscale group level coordination, simultaneously with the temporal emergence of collective processes over time (Figure 1b iii).

If macroscale movements and shape changes are linked to the individual constituents, we reasoned that individuals and groups should exhibit coordinated behavior. To test this hypothesis, we followed individual, meso- and macro-scale reorganization as a parental cell grew and separated into daughter cells (Figure 4d). To identify progression through cell division, we assessed variations in nuclear features[61] over time (Figure 4d i). Using PINE, we observed a macroscale expansion-contraction behavior consistent with literature[62], where the cell area of parental cells initially expanded corresponding to $G_1$, S and $G_2$ phases (corresponding to decreased connectivity in the model); thereafter, cell area contracted corresponding to M phase (corresponding to increased connectivity in the model), and then expanded as parental cells divided into daughter cells (corresponding to decreased connectivity in the model). Shape changes are known to be related to the cytoskeleton[63]; however, how individual constituents contribute to macroscale reorganization remain incompletely understood. Using PINE, we observed the sub-10 nm width of individual filaments remained consistent over time, indicating actin maintained as individual filaments (Figure 4d iii). By following hundreds of individual constituents (904 filaments), we discovered individual filaments also underwent expansion-contraction behavior at the individual level synchronized with macroscale shape changes: i. length of individual filaments initially contracted during $G_1$, S and $G_2$ phases (corresponding to decreased connectivity in the model). ii. next, length of individual filaments expanded during the M phase (corresponding to increased connectivity in the model). iii. finally, the length of individual filaments contracted as parental cells divided into daughter cells (corresponding to decreased connectivity in the model). At the mesoscale, the density of individual filaments also exhibited expansion-contraction behavior observed by PINE coordinated with macroscale shape changes. During $G_1$, S and $G_2$ phases, the density of individual filaments decreased (corresponding to decreased connectivity in the model). In the M phase, the density of individual filaments increased (corresponding to increased connectivity in the model). Finally, the density of individual filaments decreased as parental cells divided into daughter cells (corresponding to decreased connectivity in the model). No expansion-contraction behavior was observed in the undivided control (Figure 4e ii). Taken together, PINE revealed emergent dynamics in which individuals and groups exhibited synchronized reorganization at the individual, meso- and macro-scale levels (Figure 4e ii).

**Discussion**

In discussion, the method introduced here can be used to overcome time limitations of fluorescence super-resolution for long-time nanoscopy. Using nonbleaching PINE, we demonstrated dynamic nanoscopy of living matter over 250 hours, more than two orders of magnitude outperforming state-of-the-art fluorescence super-resolution. We resolved sub-10 nm cellular architectures and followed the emergence of their coordinated reorganization during the cell division process. We introduced a model based on local interactions of individuals to describe how individual constituents can synchronize with group level movements and shape changes at the macroscale. Presented here, we applied the model to the cell division process. This simple model provides a direct link between individual constituents and overall shape changes at the macroscale. The mechanism of coordination we proposed here in the model showed local interactions between neighboring individual constituents are sufficient to emerge into group level reorganization (Figure 4e). Analogous to disruption of animal herds under attack[64], how individual-group coordinated reorganization may be disrupted in disease are possibilities to be explored in the future. This is important to our understanding of dynamical emergent processes involved in cellular reorganization, self-organization and pattern formation. This model can be applied to other soft matter, such as shape changing materials[65].

In this work, we achieved PINE-resolved images with nanoprobe densities of $\sim 10^2/\mu m^2$, whereas higher nanoprobe densities are typically utilized in traditional fluorescence super-resolution. Since the distribution of distances from every nanoprobe to its first neighbor follows a distribution function[44], underlying protein architectures can be identified for nanoprobe labeling densities of $\sim 10^2/\mu m^2$, albeit lower than nanoprobe densities typically used in traditional fluorescence super-resolution. This offers an advantage for live imaging to minimize excessive nanoprobe labeling which may affect observed dynamics.

PINE has the potential for *in vivo* nanoscopy. A current limitation is the nanoprobe size for sufficient scattering. In the future, *in vivo* sub-10 nm nanoprobes displaying geometric singularities for extremely high field generation can be designed to be modulated by phase-intensity to overcome this limitation with interferometry. PINE has the potential for four-dimensional (4-D) nanoscopy (*t, x, y, z*). A current limitation is the sample depth and background scattering. In the future, volume (sample depth) can be achieved by employing PINE with light-sheets (*i.e.*, optical z sectioning) and background subtraction algorithms. This could lead to exciting studies of long timescale processes, such as emergent processes, evolutionary processes, ageing and age-related phenomena. New control methods, such as subdiffraction optical tweezers, used in conjunction with PINE, would create exciting possibilities for spatiotemporal control of *in vivo* processes. In conclusion, we believe PINE will open new nanoscopic opportunities for investigations demanding long-time observation windows.

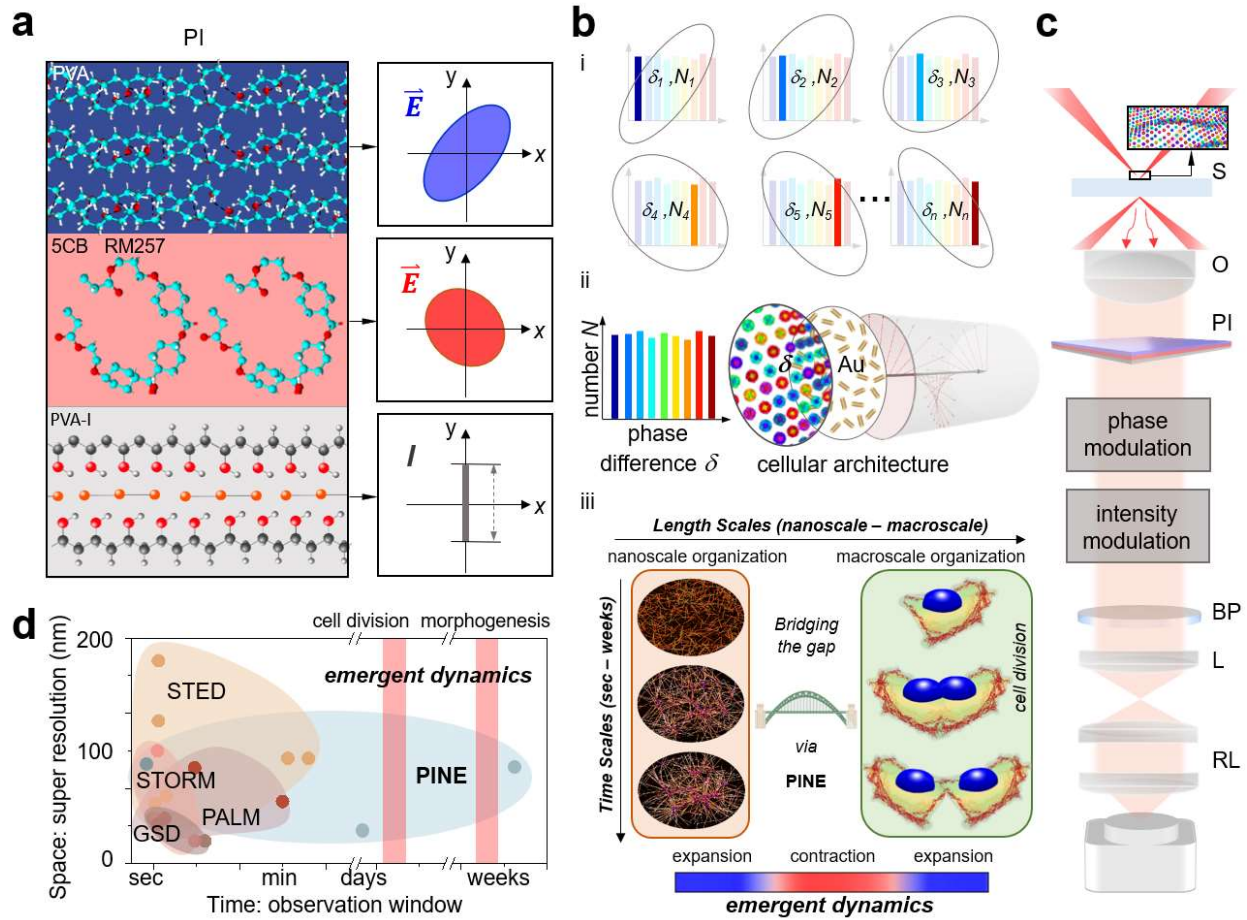

**Figure 1. Principle of the nonbleaching <u>p</u>hase <u>i</u>ntensity <u>n</u>anoscop<u>e</u>: PINE for living matter.**

**(a) Phase-intensity PI:** Integrated phase-intensity multilayer thin film consisting of polyvinyl alcohol/liquid crystalline polymers, enable precise control of phase differences between electric field components. Scattered light is reshaped according to phase modulation. Phase modulation is then converted to intensity modulation such that the resulting variation of intensity corresponds to subsets of nanoprobes labeling cellular architectures. **(b) Concept of PINE: i.** PI precisely modulates phase differences $\delta_n$ corresponding to subsets of nanoprobes $N_n$ within the population. *N*: number of nanoprobes. $\delta$: phase difference between electric field components. **ii.** Randomly distributed nanoprobes (Au nanorods) form patterns of the underlying cellular architectures. Using PI, nanoprobes exhibit phase differences between electric field components in a stochastic manner. **iii.** PINE opens a long-time investigation window to investigate emergent nanoscale-to-macroscale dynamics: in cell division, reorganization of individual constituents at the nanoscale emerges into group-level movements and shape changes at the macroscale over time. **(c) Set-up.** Darkfield configuration illuminates a nanoprobe-labeled live cell sample (S) in a temperature- and gas- controlled flow chamber. The collected scattered light by objective (O) is phase-intensity modulated (PI) and bandpass filtered (BP). To increase the system's magnification, relay lenses (RL) were added to increase the effective focal length of the tube lens (L). After phase-intensity separation, the resulting intensity variation corresponds to subsets of nanoprobes. **(d)** Fluorescence super-resolution methods, such as ground state depletion (GSD), stimulated emission depletion (STED), photo-activated localization microscopy (PALM), and stochastic optical reconstruction microscopy (STORM), have pushed spatial resolution beyond the diffraction limit (y-axis). PINE creates new nanoscopic opportunities along the time axis (x-axis) for **investigations demanding long-time observation windows.**

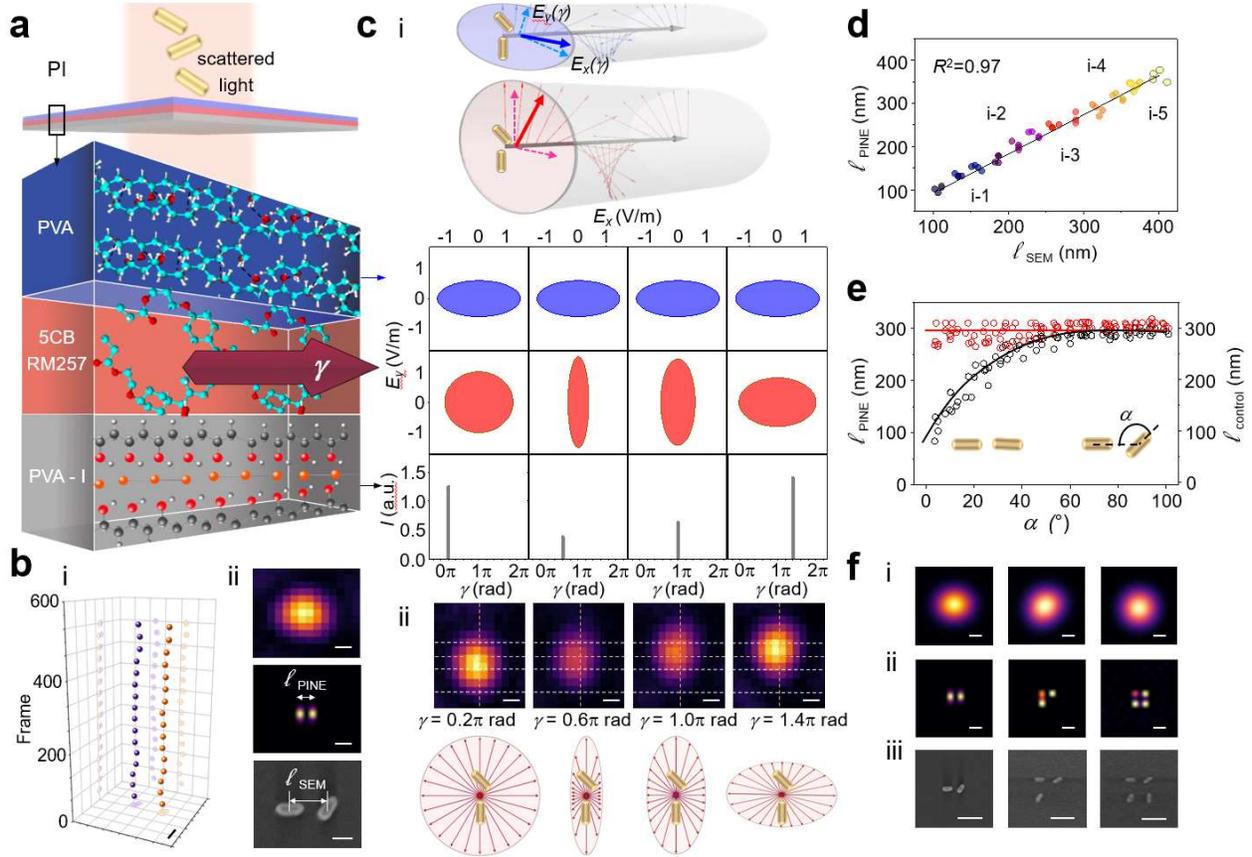

**Figure 2. PI phase-intensity separates multiple nanoprobes within a diffraction limited region by distinguishing phase differences between electric field components. (a) PI:** Integrated phase-intensity multilayer stack of polyvinyl alcohol, liquid crystal and liquid crystal polymer, and polyvinyl alcohol-iodine layers. **(b) Phase-intensity separation: i)** Scatter plot: Images repeatedly acquired over the modulation range $\gamma$ from 0 to $2\pi$ rad, generating image stack of ~500 total frames. The spatial positions of each nanorod were determined by taking the median value. Scale bar: 100 nm. **ii)** Diffraction limited darkfield image. Scale bar: 200 nm. PINE-resolved image. Color represents intensity after phase-to-intensity conversion. Scale bar: 200 nm. Scanning electron microscopy image. Scale bar: 100 nm. **(c) i)** Schematic of multiple nanoprobes (gold nanorods) within a diffraction limited region: Reshaped scattered light exhibited phase difference between electric field components $E_x$ and $E_y$ to distinguish multiple nanoprobes within a diffraction limited region. Calculated electric field and intensity amplitudes: Scattered light was

reshaped (top row), phase modulated (middle row), and intensity modulated (bottom row). **ii)** Experimental darkfield images at $\gamma$ = 0.2π, 0.6π, 1.0π and 1.4π rad acquired by PI. Scale bar: 200 nm. Schematic of phase modulation by PI. **(d) Benchmarking of localization:** Comparison of $\ell_{PINE}$ imaged by PINE and $\ell_{SEM}$ measured by scanning electron microscopy. **(e) Localization precision:** Red curve $\ell_{PINE}$: Displacement-free PINE showing precise localization as $\alpha \to 0$. Black curve $\ell_{control}$: Displacement control showing increasingly imprecise localization as $\alpha \to 0$. **(f) i)** Diffraction-limited darkfield images of multiple nanoprobes [two (left), three (middle), four (right)]. Scale bar: Scale bar: 200 nm. **ii)** PINE-resolved images of multiple nanoprobes [two (left), three (middle), four (right)]. Color represents intensity after phase-to-intensity conversion. Scale bar: 200 nm. **iii)** Scanning electron microscopy images of multiple nanoprobes [two (left), three (middle), four (right)]. Scale bar: 250 nm.

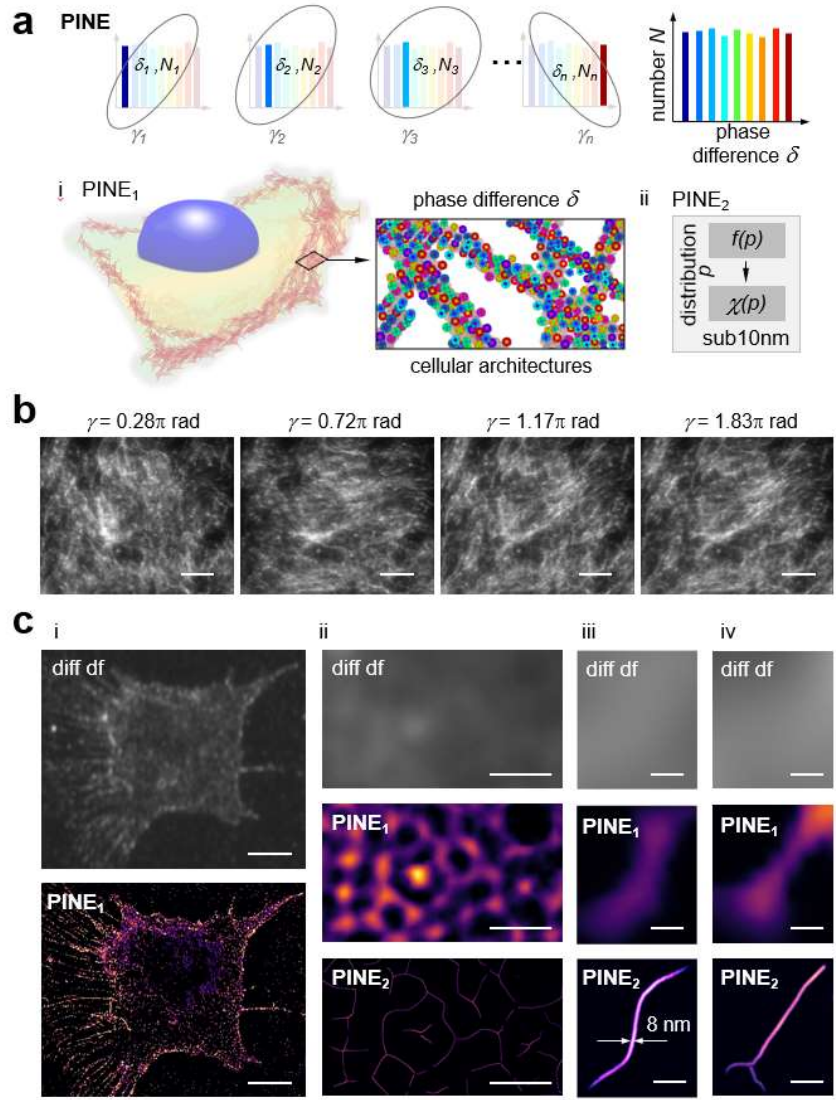

**Figure 3. PINE reveals sub-10 nm cellular architectures in SH-SY5Y cells. (a) PINE: i)** PINE$_1$: PI precisely modulates phase differences $\delta_n$ corresponding to subsets of nanoprobes $N_n$ within the population. *N*: number of nanoprobes (gold nanorods). $\delta$: phase difference between electric field components. In cells, randomly distributed nanoprobes (gold nanorods) form patterns of the underlying cellular architectures. Using PI, nanoprobes exhibit phase differences between electric field components in a stochastic manner. **ii)** PINE$_2$: Surface or curvilinear features *f(p)* were defined. Parameter $\chi$ was obtained from the pattern of precisely localized nanoprobe populations and *f(p)*. From the distribution $\chi(p)$, sub-10 nm information was obtained. **(b)** Experimental

darkfield images of cellular architectures (actin) at $\gamma = 0.28\pi$, $0.72\pi$, $1.17\pi$ and $1.83\pi$ rad acquired using PI. Scale bar: 20 μm. **(c) i)** (top) Diffraction-limited darkfield image of SH-SY5Y cell. (bottom) PINE$_1$-resolved image of SH-SY5Y cell. Scale bar: 10 μm. **ii)** (top) Diffraction-limited darkfield image. (middle) PINE$_1$-resolved image. (bottom) PINE$_2$-resolved image. Scale bar: 500 nm. **iii) PINE revealed sub-10 nm actin filament architecture in SH-SY5Y cells:** (top) Diffraction-limited darkfield image. (middle) PINE$_1$-resolved image. (bottom) PINE$_2$-resolved image. Scale bar: 80 nm. **iv) PINE revealed sub-10 nm branched actin architecture in SH-SY5Y cells:** (top) Diffraction-limited darkfield image. (middle) PINE$_1$-resolved image. (bottom) PINE$_2$-resolved image. Scale bar: 80 nm.

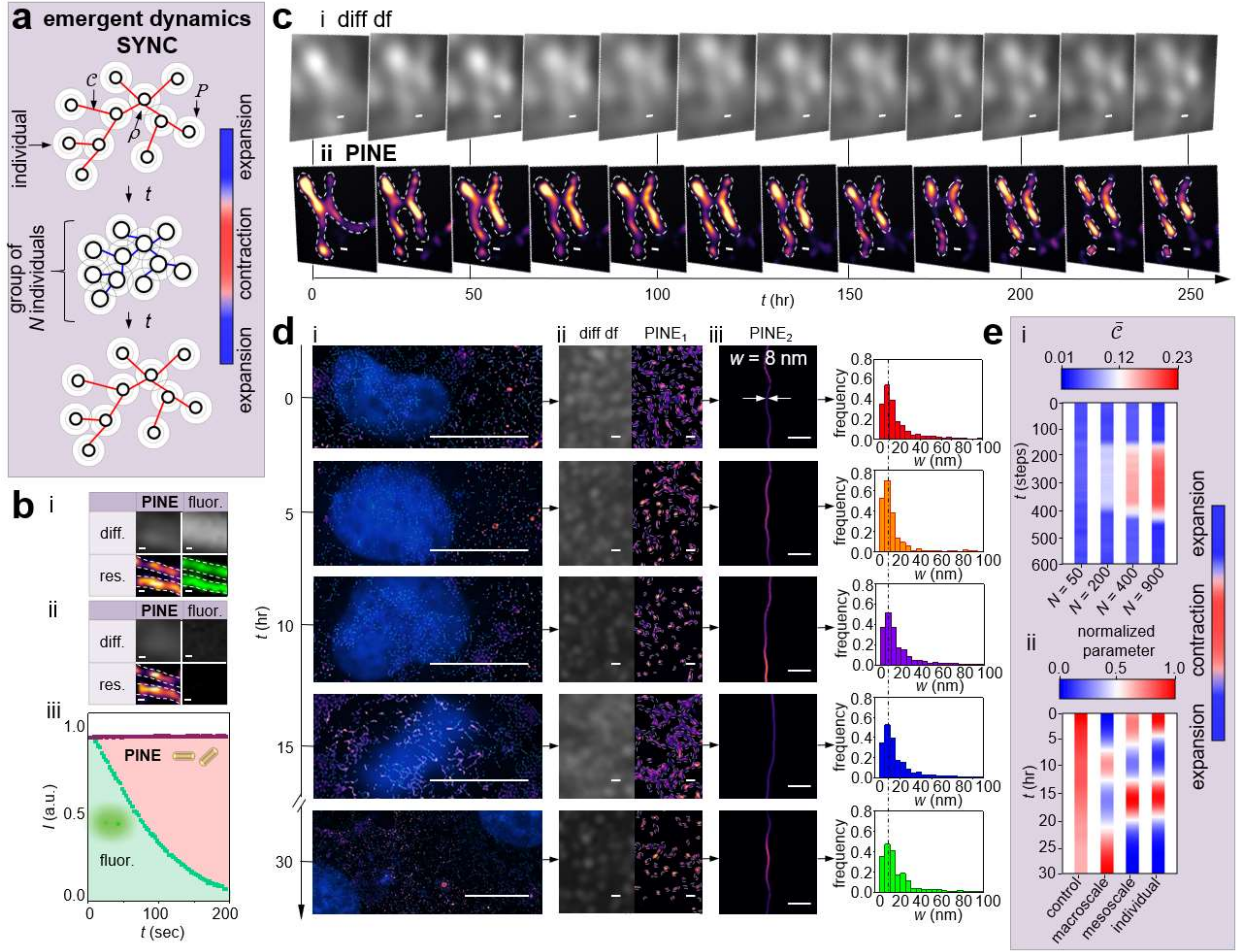

**Figure 4. <u>S</u>ynchronized i<u>n</u>dividual-group <u>c</u>ontraction-expansion, SYNC, emerged during the cell division process. (a) SYNC Model:** Conceptual schematic of SYNC model following Eq.4 - 6 where $t$ = time, $P$ = long-range space, $\rho$ = short-range space, $(1 - \mathcal{C})$ = connection. **(b) Benchmarking of nonbleaching nanoscopy:** fluorescence super-resolution versus PINE. In PINE, labeling nanoprobes' nonbleaching role enables long-time nanoscopy to follow long-time processes. **i)** Time $t$ = 0 sec: Diffraction-limited darkfield image at time $t$ = 0 sec (top left) and corresponding PINE-resolved image at time $t$ = 0 sec (bottom left). Diffraction-limited fluorescence image at time $t$ = 0 sec (top right) and corresponding fluorescence super-resolution resolved image at time $t$ = 0 sec (bottom right). Scale bar: 200 nm. **ii)** Time $t$ = 200 sec: Diffraction-limited darkfield image at time $t$ = 200 sec (top left) and corresponding PINE-resolved image at time $t$ = 200 sec (bottom left). Diffraction-limited fluorescence image at time $t$ = 200 sec (top right) and

corresponding fluorescence super-resolution resolved image at time $t$ = 200 sec (bottom right). Scale bar: 200 nm. **iii)** Graph of average scattering intensity over time for fluorescence super-resolution versus PINE. PINE (purple), fluorescence super-resolution (green). **(c) Validation of long-time PINE:** Diffraction-limited darkfield time course of actin dynamics. PINE-resolved time course of actin dynamics. Scale bar: 150 nm. **(d) i) Cell division process at the macro-, meso- and individual levels:** $PINE_1$-resolved macroscale time course in SH-SY5Y cells (left) overlaid with fluorescence (nucleus) to assess nuclear features. Scale bar: 10 μm. **ii)** Diffraction-limited darkfield images and $PINE_1$-resolved time course of mesoscale group of individual filaments in SH-SY5Y cells (middle). Scale bar: 500 nm. **iii)** $PINE_2$-resolved time course of individual filaments in SH-SY5Y cells (right). Scale bar: 80 nm. Time course histograms of resolved actin filament width by PINE. **(e) SYNC: i) Theory: Model applied to cell division.** Output connection $(1 - \mathcal{C})$ over time for different group sizes of $N$ individuals according to Eq.4 - 6. $\bar{\mathcal{C}}$: mean $\mathcal{C}$ for a group size of $N$ individuals. As the group size became larger, the expansion-contraction behavior increased, $c$ = 0.5 and 5, $P$ = 5 μm, $\rho$ = 500 nm. **ii) Experiment: Synchronized expansion-contraction at the macro-, meso- and individual levels**. Time course contraction-expansion graph was normalized in order to plot together undivided control, macroscale, mesoscale, individual. Number of individuals $N$ = 904.

**Acknowledgements**

This work was supported by the Air Force Office of Scientific Research (AFOSR FA9550-16-1-0272, FA9550-19-1-0186, FA9550-22-1-0285), National Science Foundation (NSF 1454188), and academic research fund at the University of Michigan.